\documentclass[11pt,a4paper]{article}
\usepackage{jheppub}
\usepackage{gensymb}
\usepackage{placeins}
\usepackage{tabularx}
\usepackage{array,booktabs,longtable,tabularx}
\newcolumntype{L}[1]{>{\raggedright\arraybackslash}p{#1}}
\usepackage{subfigure}
\usepackage{amssymb,amsmath}
\usepackage{graphicx}
\usepackage{color}
\usepackage{cancel}
\usepackage{setspace}
\usepackage{makecell}

\def\be{\begin{equation}}
\def\ee{\end{equation}}
\def\bea{\begin{eqnarray}}
\def\eea{\end{eqnarray}}
\def\bi{\begin{itemize}}
\def\ei{\end{itemize}}
\begin{document}

\begin{titlepage}
\pagestyle{empty}

\vspace*{0.2in}
\begin{center}
{\Large \bf  Mono-$\mathbf{\it Z^{\prime}}$ Signatures in the $B\!-\!L$  \\[0.25cm] Supersymmetric Standard Model at the LHC } \\[0.2cm]
%{\Large \bf Exploring Mono $\mathbf{\it Z^{\prime}}$ Signatures at the LHC\\[0.25cm]
 %to Elucidate the Spin Properties of Dark Matter } \\
\vspace{1cm}
{\large Ali \c{C}i\c{c}i$^{a,b,}$\footnote{E-mail: ali.cici@cern.ch}, Shaaban Khalil$^{c,}$\footnote{E-mail: skhalil@zewailcity.edu.eg}, Stefano Moretti$^{d,e,}$\footnote{E-mail: s.moretti@soton.ac.uk;~stefano.moretti@physics.uu.se} and 
Cem Salih $\ddot{\rm U}$n$^{f,}$\footnote{E-mail: cemsalihun@uludag.edu.tr}}
\vspace{0.5cm}

{\small \it

$^{a}$Orhaneli T\"{u}rkan-Sait Y\i lmaz High School, TR16980 Bursa,T$\ddot{u}$rkiye
\\  $^b$Department of Physics, Faculty of Engineering and Natural Sciences, \\ Bursa Technical University, TR16310, Bursa, T\"{u}rkiye

$^c$Center for Fundamental Physics, Zewail City of Science and Technology, 6 October City, Giza 12588, Egypt\\
$^d$School of Physics and Astronomy, University of Southampton, Southampton, SO17 1BJ,\\ United Kingdom \\
$^e$Department of Physics and Astronomy, Uppsala University, Box 516, SE-751 20 Uppsala, Sweden\\
$^f$Department of Physics, Bursa Uluda\~{g} University, TR16059 Bursa, Turkey}

\end{center}

\setcounter{footnote}{0}

\vspace{0.5cm}
\begin{abstract}
\; 
The $B-L$ Supersymmetric Standard Model with Inverse Seesaw (BLSSM-IS) extends the Minimal Supersymmetric Standard Model (MSSM) by incorporating a gauged $B\!-\!L$ symmetry, right-handed neutrinos and an additional neutral gauge boson $Z^{\prime}$. Searches at the Large Hadron Collider (LHC) constrain the mass of this gauge boson to be as low as only $\approx2.3~\text{TeV}$ in the BLSSM-IS, owing to
interference effects with the SM. In this framework, mono-$Z^{\prime}$ events can arise from the associated production of a $Z^{\prime}$ boson and a singlet Higgs boson $h^{\prime}$, where $h^{\prime}$ subsequently decays into missing energy carried by a pair of the Lightest Supersymmetric Particle (LSP)---either a neutralino or a right-handed sneutrino---which serves as a Dark Matter (DM) candidate. Focusing on leptonic decays of the $Z^{\prime}$ (electrons and muons), we analyse the kinematic distributions of the final-state leptons and the missing transverse energy in order to extract a signal for this process which is independent of the nature of the BLSSM-IS DM.

\end{abstract}
\end{titlepage}

\section{Introduction}
\label{sec:intro}

Despite its impressive success, the Standard Model (SM) undeniably faces certain limitations and shortcomings, such as the gauge hierarchy problem, the lack of suitable Dark Matter (DM) candidates and the absence of neutrino masses and mixing. Addressing each of these issues requires extensions to the SM, involving modifications of its gauge group and/or particle content. 
In this context, Supersymmetry (SUSY) stands out as one of the leading candidates for extending the SM. SUSY offers solutions to various problems encountered by the SM, such as stabilizing the Higgs boson mass against quadratic divergences and providing numerous DM candidates when R-parity is conserved.
However, while the Minimal Supersymmetric SM (MSSM) maintains the integrity of the SM gauge group but it falls short in accommodating neutrino masses and mixing that are consistent with experimental observations  \cite{Himmel:2015cna,Super-Kamiokande:2014ndf,ParticleDataGroup:2024cfk}.

Neutrinos offer the potential for yet another extension of the MSSM \cite{Moretti:2019ulc}, known as the $B-L$ Supersymmetric Standard Model (BLSSM), which supplements the MSSM gauge group with a gauged $B-L$ symmetry. In this extended model, the entire symmetry at a high scale is $SU(3)_c \times SU(2)_L \times U(1)_Y \times U(1)_{B-L}$, the $B-L$ gauge group undergoes spontaneous breaking facilitated by two chiral MSSM singlet fields, $\chi_{1,2}$, with $B-L$ charges assigned as $\pm 1$. This extension of the MSSM finds strong motivation in its ability to gauge the accidental $B-L$ symmetry of the SM and necessitates the inclusion of three right-handed neutrinos for anomaly cancellations. Consequently, $B-L$ extensions of the MSSM provide a suitable framework for implementing the Seesaw mechanisms.
While the Type I Seesaw mechanism represents one of the simplest implementations for generating neutrino masses and mixing, it either demands very heavy right-handed neutrinos or requires finely tuned small Dirac Yukawa couplings between the Higgs boson and neutrinos $(Y_{\nu} \lesssim 10^{-7})$ \cite{Pandey:2025uah}. As a result, the right-handed neutrinos and their superpartners decouple from the MSSM particles at a high scale, leaving the phenomenology of the MSSM largely intact \cite{Khalil:2006yi,Basso:2008iv,Basso:2010yz,Majee:2010ar,Li:2010rb,FileviezPerez:2009hdc,DelleRose:2017ukx,Un:2016hji,Jones-Perez:2023med,Gupta:2025vcp}. 

On the contrary, adopting the Inverse Seesaw (IS) mechanism (hereafter, we use BLSSM-IS to refer to the ensuing construct) instead of Type I introduces a more active role for neutrinos in phenomenology. In this scenario, the generation of small neutrino masses does not necessitate tiny neutrino Yukawa couplings, even when the right-handed neutrinos and their superpartners appear in the spectrum at sub-TeV scales. Consequently, the presence of these neutrinos (and sneutrinos) can significantly influence various phenomenological aspects, such as DM dynamics \cite{Abdallah:2017gde,El-Zant:2013nta}, contributions to the muon anomalous magnetic moment  \cite{Khalil:2015wua,Altin:2017sxx,Abdallah:2011ew} and deviations in \(B-\)meson decays that align with current experimental data \cite{Boubaa:2022xsk}. These factors collectively motivate a thorough exploration of the implications of the IS mechanism.

In addition to the relevant experimental tests in the aforementioned contexts, $B-L$ extensions are directly probed by the Large Hadron Collider (LHC) multi-purpose experiments, ATLAS and CMS, as they have both been pursuing neutral gauge boson signals associated with the $U(1)_{B-L}$ group. Furthermore, beyond such direct searches at the  LHC, the $B-L$ sector can manifest itself in numerous other processes due to its substantial extension of the particle content with, e.g.,  new MSSM singlet Higgs bosons. These Higgs bosons may even have lower masses than the SM-like Higgs boson, potentially causing anomalies  in $\phi \rightarrow \gamma\gamma$ at light mass scales  \cite{Abdelalim:2020xfk,Diaz:2024yfu} or in events involving mono-Higgs \cite{Abdallah:2016vcn} and mono-photon signatures \cite{Hicyilmaz:2023tnr}.

\begin{figure}[h!]
\centering
\includegraphics[scale=0.4]{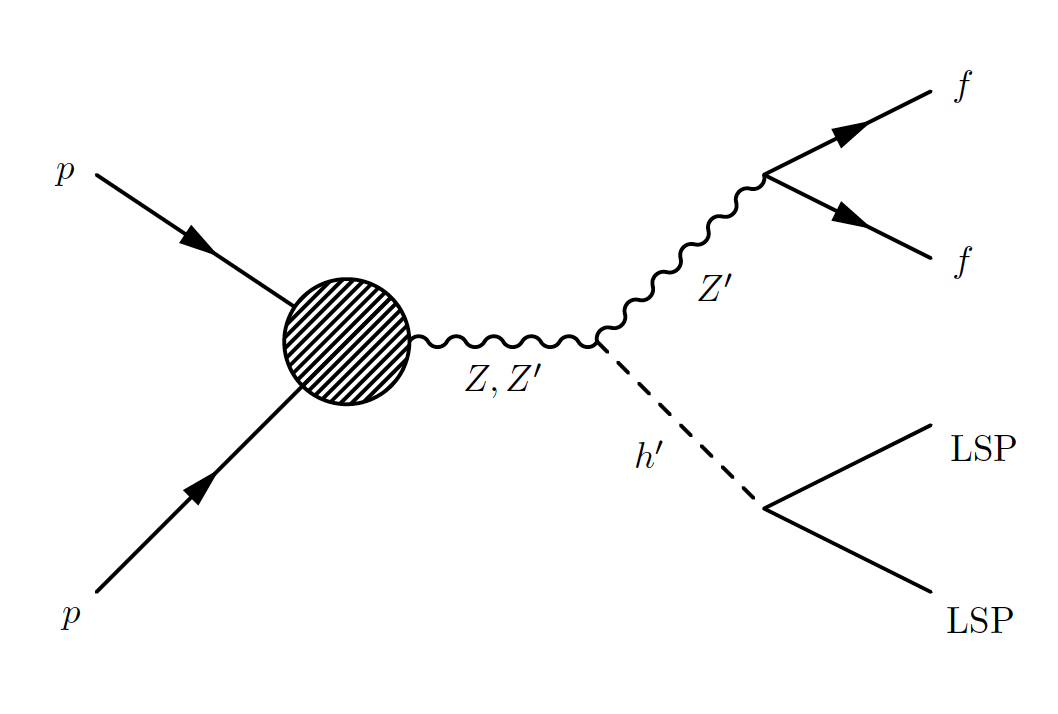}
\caption{Representative diagram for mono-$Z^{\prime}$ events. Here, $f= l~(e,\mu)$ and LSP $={\tilde \chi}_1^0$,  ${\tilde \nu}_R$. }
\label{fig:signal}
\end{figure}

In this study, we will investigate mono-$Z^{\prime}$ events, wherein a $Z^{\prime}$ boson is produced in association with a singlet Higgs boson ($h^{\prime}$) via $s$-channel off-shell  production of a $Z$  or  $Z^{\prime}$ boson. The final states of these processes consist of quarks or leptons from $Z^{\prime}$ decays and missing energy resulting from $h^{\prime} \rightarrow {\rm LSP~LSP}$. The stable LSP in the  BLSSM-IS could be, e.g.,  either the lightest neutralino (representative of a fermionic type of DM) or the lightest right-handed sneutrino (representative of  a bosonic type of DM) \cite{DelleRose:2017uas}. A schematic representation of such processes is depicted in Figure \ref{fig:signal}. While hadronic decays of $Z^{\prime}$ can be significant, we will focus on the leptonic ones as they offer a cleaner signal with less background and thus lead to more sensitive results. 

The paper is organized as follows.  Section 2 provides a brief review of the BLSSM-IS, emphasising the masses and interactions of the $Z'$ boson, the lightest neutralino and the lightest right-handed sneutrino. Section 3 discusses the scanning procedure and constraints used. Section 4 dwells on the mono-$Z'$ events generated in the BLSSM-IS at the LHC, in terms of relevant cross-sections, given some LSP, $h'$ and $Z'$ masses. In Section 5, we concentrate on event selection to extract the signal from the SM background. Final remarks and conclusions are presented in Section 6.

\section{BLSSM-IS: Model Structure and LSP}
\label{sec:model}

The BLSSM-IS extends the MSSM gauge group to $SU(3)_C\times SU(2)_L\times U(1)_Y\times U(1)_{B-L}$, where the $U(1)_{B-L}$ symmetry is spontaneously broken by the Vacuum Expectation Values (VEVs) of the chiral singlet Superfields $\hat{\eta}_{1,2}$ with $B-L$ charges $\pm 1$. The TeV-scale BLSSM-IS refers to the case in which the $U(1)_{B-L}$ symmetry can be broken at $\mathcal{O}({\rm TeV})$. Supplementing the model with the IS mechanism and ensuring anomaly cancellations require particle spectrum to include three generations of chiral singlets $\hat{\nu}^c_i$ ($B-L = -1$), $\hat{S}_1$ ($B-L = +2$), and $\hat{S}_2$ ($B-L = -2$), along with a $Z'$ gauge boson. The Superpotential is given by \cite{Hirsch:2009ra,Khalil:2010iu}

\be
W =  Y_u\hat{Q}\hat{H}_2\hat{U}^c + Y_d \hat{Q}\hat{H}_1\hat{D}^c + Y_e\hat{L}\hat{H}_1\hat{E}^c + Y_\nu\hat{L}\hat{H}_2\hat{\nu}^c + Y_S\hat{\nu}^c\hat{\eta}_1\hat{S}_2 + \mu\hat{H}_1\hat{H}_2 + \mu'\hat{\eta}_1\hat{\eta}_2.
\label{superpotential}
\ee

The symmetry breaking leads to the physical masses for the neutral gauge bosons as follows:

\begin{equation}
M_Z^2 = \frac{1}{4}(g_1^2 + g_2^2)v^2, \qquad 
M_{Z'}^2 = g_{BL}^2 v'^2 + \frac{1}{4} \tilde{g}^2 v^2,
\end{equation}
where \( v = \sqrt{v_1^2 + v_2^2} \simeq 246~\text{GeV} \) is the Electro-Weak (EW) VEV  and \( v' = \sqrt{v_1'^2 + v_2'^2}\) determines the scale of \( B\!-\!L \) symmetry breaking.

The neutralino sector is also significantly extended such that the physical spectrum involves seven neutralinos \( \tilde{\chi}^0_i \) (\( i = 1, \dots, 7 \)), which are mass eigenstates formed from linear combinations of the gauginos (\( \tilde{B}, \tilde{W}^3, \tilde{B}' \)), the MSSM Higgsinos (\( \tilde{H}_1^0, \tilde{H}_2^0 \)) and the \( B-L \) Higgsinos (\( \tilde{\eta}_1, \tilde{\eta}_2 \)) as

\be
\tilde{\chi}^0_1 = V_{11} \tilde{B} + V_{12} \tilde{W}^3 + V_{13} \tilde{H}_1^0 + V_{14} \tilde{H}_2^0 + V_{15} \tilde{B}' + V_{16} \tilde{\eta}_1 + V_{17} \tilde{\eta}_2.
\ee
The lightest neutralino (LSP) is referred to as pure \( \tilde{B}' \) if \( V_{15} \simeq 1 \) and \( V_{1i} \simeq 0 \) for \( i \neq 5 \) or as pure \( B-L \) Higgsino if \( V_{16} \) or \( V_{17} \simeq 1 \). 

Now we turn to the sneutrino mass matrix. We define the sneutrino fields as:
\be
\tilde{\nu}_L = \frac{1}{\sqrt{2}} (\tilde{\nu}_L^+ + i\tilde{\nu}_L^-), \quad \tilde{\nu}_R = \frac{1}{\sqrt{2}} (\tilde{\nu}_R^+ + i\tilde{\nu}_R^-), \quad \tilde{S}_2 = \frac{1}{\sqrt{2}} (\tilde{S}_2^+ + i\tilde{S}_2^-).
\ee
The sneutrino mass matrix can then be written as:
\be
M_{\nu}^2 = \begin{pmatrix} 
{\cal M}^2_{\rm even} & 0 \\
0 & {\cal M}^2_{\rm odd}
\end{pmatrix},
\ee
where the CP-even and CP-odd sneutrino mass matrices are given in \cite{Khalil:2015naa}.
The diagonalization of the mass matrix, particularly with a non-zero gluino mass, is a numerically challenging task. The lightest CP-even and CP-odd sneutrinos, \( \tilde{\nu}^+_i \) and \( \tilde{\nu}^-_i \), are nearly degenerate in mass and the heavier sneutrinos arise from mixing between \( \tilde{\nu}_R \) and \( \tilde{S}_2 \). The off-diagonal elements of \( {\cal M}^2_{\pm} \) are small, allowing for an approximate block-diagonal form. For typical values of \( \mu' \) and/or \( A_S \sim {\cal O}(1) \text{ TeV} \), one eigenvalue of the matrix is light, with the lightest sneutrino mass of order \( 100 \text{ GeV} \). 

The lightest sneutrino, \( \tilde{\nu}_1 \), can be written as a linear combination of CP-even sneutrino fields:
\be
\tilde{\nu}_1 = \sum_{i=1}^3 R_{1i} \tilde{\nu}^+_L + \sum_{j=4}^6 R_{1j} \tilde{\nu}^+_R + \sum_{k=7}^9 R_{1k} \tilde{S}^+_2.
\ee
For typical soft SUSY-breaking parameters, the coefficients are approximately 
\[
R_{1l} = \frac{1}{\sqrt{2}} \{0, 0, 0, 1, 0, 0, 1, 0, 0\},
\]
implying that \( \tilde{\nu}_1 \) is predominantly right-handed, composed mainly of \( \tilde{\nu}^+_R \) and \( \tilde{S}^+_2 \).
In \cite{Abdallah:2017gde}, it was demonstrated that the additional \( B-L \) neutralinos —namely the $B'$ gaugino and the two extra higgsinos  \( \tilde{\eta}_1, \tilde{\eta}_2 \))— as well as the lightest right-handed sneutrino can serve as the LSP and constitute viable DM candidates.

%%%%%%%%%%%%%%%%%%%%%%%%%%%%%%%%%%%%%%
\section{Scanning Procedure and Experimental Constraints}
\label{sec:scan}

In this section, we briefly describe the data generation and phenomenological analyses confronting theoretical solutions with current experimental results. We use SPheno-4.0.4 \cite{Porod:2003um, Goodsell:2014bna} to compute the mass spectrum and Branching Ratios (BRs), which are relevant to the signal events represented diagrammatically in Figure \ref{fig:signal}. We allow almost all low-scale parameters to vary by scanning over masses, the $U(1)_{B-L}$ gauge coupling and the gauge kinetic mixing between $U(1)_{B-L}$ and $U(1)_{Y}$, parameterized by $g^{\prime}$. The neutrino Yukawa matrix is determined by adopting the Casas-Ibarra parametrization \cite{Casas:2001sr}, suitably modified for the IS mechanism \cite{Hammad:2016bng} to ensure consistency with experimental neutrino data \cite{Himmel:2015cna,Super-Kamiokande:2014ndf,Lokhov:2022zfn,ParticleDataGroup:2024cfk}. The scanning procedure employs the Metropolis-Hastings algorithm \cite{Baer:2008jn, Belanger:2009ti}, accepting only solutions where the neutral LSP is suitable for producing missing energy in the final state. After calculating the mass spectrum and BRs, we use CalcHEP \cite{Belyaev:2012qa} to compute the inclusive cross-sections for each solution and MadGraph \cite{Alwall:2014bza,Alwall:2014hca} for detailed differential analyses based on specific Benchmark Points (BPs). Consistency between the results from CalcHEP and MadGraph is verified through rigorous cross-checking.

After generating the parameter space points, the solutions are subjected to several constraints including the mass bounds on the Supersymmetric particles \cite{ParticleDataGroup:2014cgo,ATLAS:2021twp,ATLAS:2020syg,ATLAS:2022rcw} and Higgs bosons \cite{ATLAS:2012yve,CMS:2012qbp,CMS:2013btf} as well as constraints from combined results for rare $B-$meson decays \cite{CMS:2020rox,Belle-II:2022hys,HFLAV:2022esi}. Among the mass bounds, we apply the current limits on the gluino ($\gtrsim 2.2$ TeV) and squark ($\gtrsim 1.5$ TeV) masses \cite{ATLAS:2023afl}. These constraints significantly suppress any additional mixing in the quark flavors and quark masses are diagonalized using the Cabibbo-Kobayashi-Maskawa (CKM) matrix, as in the SM. In contrast, there are no severe bounds on the sleptons. With the presence of right-handed neutrinos, charged leptons can mix significantly, potentially leading to large Lepton-Flavor Violating
(LFV) processes. These are constrained by data from studies of $\mu \rightarrow e\gamma$, $\tau \rightarrow e\gamma$ and $\tau \rightarrow \mu\gamma$, with the strongest limits  coming from $\mu \rightarrow e\gamma$ \cite{MEG:2016leq,ParticleDataGroup:2022pth}. 

One of the main issues in our analyses is the precision and consistency in the  calculation of the SM-like Higgs boson ($h_{\rm SM}$) properties, as the latter are affected by the presence of additional (neutral and CP-even) Higgs states in
the BLSSM-IS. Despite significant improvements in the SPheno package \cite{Goodsell:2014bna}, there still remains about 2 GeV uncertainty in theoretical calculations of the Higgs boson mass due to the large mixing in the sfermion sector and uncertainties in the experimental analyses of top quark mass and strong coupling \cite{Baer:2021tta}. In addition to its mass, its couplings and decay modes are also crucial in our analyses. The constraints from rare $B-$meson provides also a fit for the SM-like Higgs couplings to quarks and leptons in almost a model independent way (for a recent discussion, see Section 3 of \cite{Nis:2025fxc}). Besides, the presence of interacting neutrinos induces invisible and LFV decays for the SM-like Higgs boson, which are constrained in our analyses with current experimental results \cite{CMS:2023sdw,ATLAS:2023mvd}. The constraints used in our analysis can be summarized as follows.

\begin{equation}
\begin{array}{rll}
& m_{\tilde{e}_{i}},m_{\tilde{\chi}^{\pm}} & \geq 105 ~{\rm GeV},\\
& m_{\tilde{g}} & \geq 2.2~{\rm TeV}, \\
& m_{\tilde{q}} & \geq 1.5~{\rm TeV}, \\
123~{\rm GeV} \leq & m_{h_{\rm SM}} & \leq 127~{\rm GeV}, \\
1.95\times 10^{-9} \leq & {\rm BR}(B_{s}\rightarrow \mu \mu) & \leq 3.43\times 10^{-9}, \\ 
2.99\times 10^{-4} \leq & {\rm BR}(B\rightarrow X_{s} \gamma) & \leq 3.87\times 10^{-4}, \\
& {\rm BR}(\mu \rightarrow e\gamma) & \leq 4.3\times 10^{-13}, \\
& {\rm BR}(\tau \rightarrow e\gamma) & \leq 3.3\times 10^{-8}, \\
& {\rm BR}(\tau \rightarrow \mu\gamma) & \leq 4.4\times 10^{-8}, \\
& {\rm BR}(h_{\rm SM} \rightarrow e\mu) & \leq 4.4\times 10^{-5}, \\
& {\rm BR}(h_{\rm SM} \rightarrow e\tau) & \leq 0.2, \\
& {\rm BR}(h_{\rm SM} \rightarrow \mu\tau) & \leq 0.15, \\
& {\rm BR}(h_{\rm SM} \rightarrow {\rm invisible}) & \leq 0.15, \\
& M_{Z^{\prime}}/g_{B-L} & \geq 7 \, \text{TeV}.
\end{array}
\label{eq:constraints}
\end{equation}

Among these constraints, the mass of the $Z^{\prime}$ also imposes a significant restriction on the parameter space. Although the sensitivity depends on the gauge coupling of $U(1)_{B-L}$, rigorous experimental analyses have already established a lower bound on $Z^{\prime}$ around 3 TeV or higher \cite{CMS:2021ctt,Scutti:2021kjp,CMS:2023ooo,CMS:2024vhy}. However, notice that such limits, obtained by the mentioned experiments under the assumption of a pure $Z'$ signal, can be substantially reduced in the BLSSM-IS, quite irrespectively of the actual neutrino sector, in presence of interference effects between the $Z'$ and the SM background, particularly because  of the purely vector couplings of a $Z'$ of $B-L$ origin, which then interferes very strongly (and negatively) with the $\gamma$ and $Z$ amplitudes in both the Drell-Yan and vector boson pair production modes \cite{Abdallah:2015uba,Preparation}. {Thus, we require $Z^{\prime}$ mass to be consistent with the LEP2 bounds by imposing $M_{Z^{\prime}}/g_{B-L} \geq 7 \, \text{TeV}$ \cite{ALEPH:2013dgf}.}

%In our analysis, we also consider the model-independent constraint on the $Z^{\prime}$ mass, as imposed by LEP2. {\textcolor{red}{Citation needed here!}}

At the end of the analysis we will select two BPs one for each case of LSP, i.e., sneutrino and neutralino, with, crucially, the mass of the two being nearly degenerate, so as not to bias the analysis through kinematic effects. 
Likewise for the $Z'$ mass: it should essentially be the same for both BPs for the same reason.
The solutions are selected such that they are consistent with all the constraints listed in Eq.~(\ref{eq:constraints}) and yield among the largest cross-sections for mono-$Z^{\prime}$ events, as shown in Figure \ref{fig:signal}. These solutions are then used for detailed detector analyses and discussed in Section \ref{Sec:SimAnalysis}. %We first generate the signal and relevant background events at the parton level with MadGraph5\_aMC@NLO (v3.6.3) \cite{Alwall:2014hca}. Then the generated events are transferred to Pythia 8.3 \cite{Sjostrand:2014zea} which handles the hadronization, parton shower and unstable particles in final states. The detector responses to the final states are also included through the use of Delphes-3.5.0 \cite{deFavereau:2013fsa}.

%%%%%%%%%%%%%%%%%%%%%%%%%%%%%%%%%%%%%%%%%%%%%%%%%%%%%%%
\section{Signal Cross-Sections, LSP and $Z^{\prime}$ Masses}

\begin{figure}[t!]
\centering
\subfigure{\includegraphics[width=6cm,height=5cm]{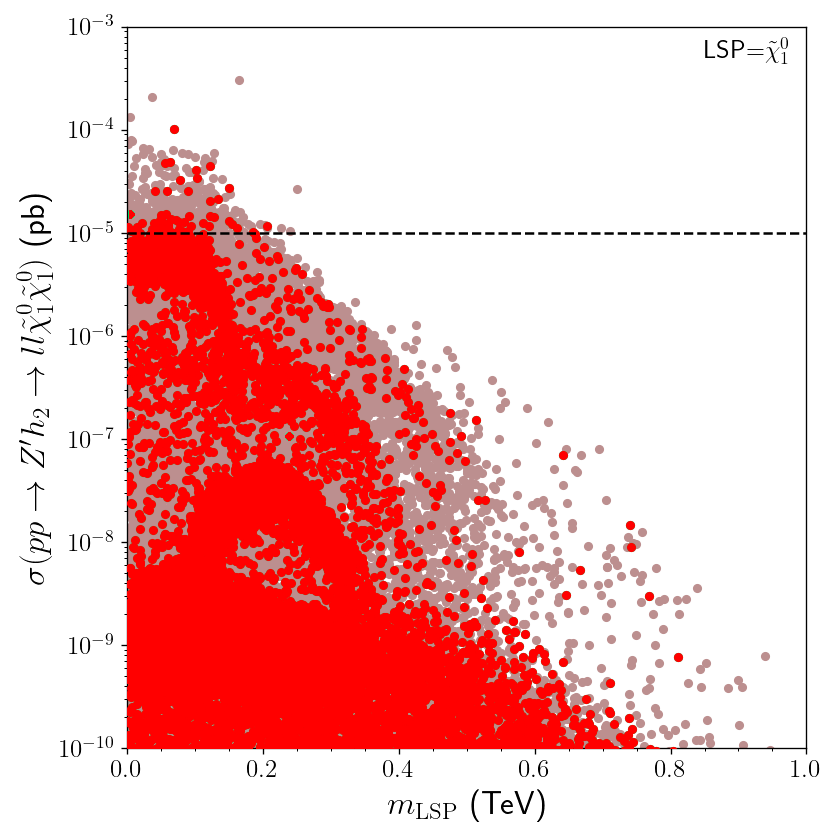}}~~~
\subfigure{\includegraphics[width=6cm,height=5cm]{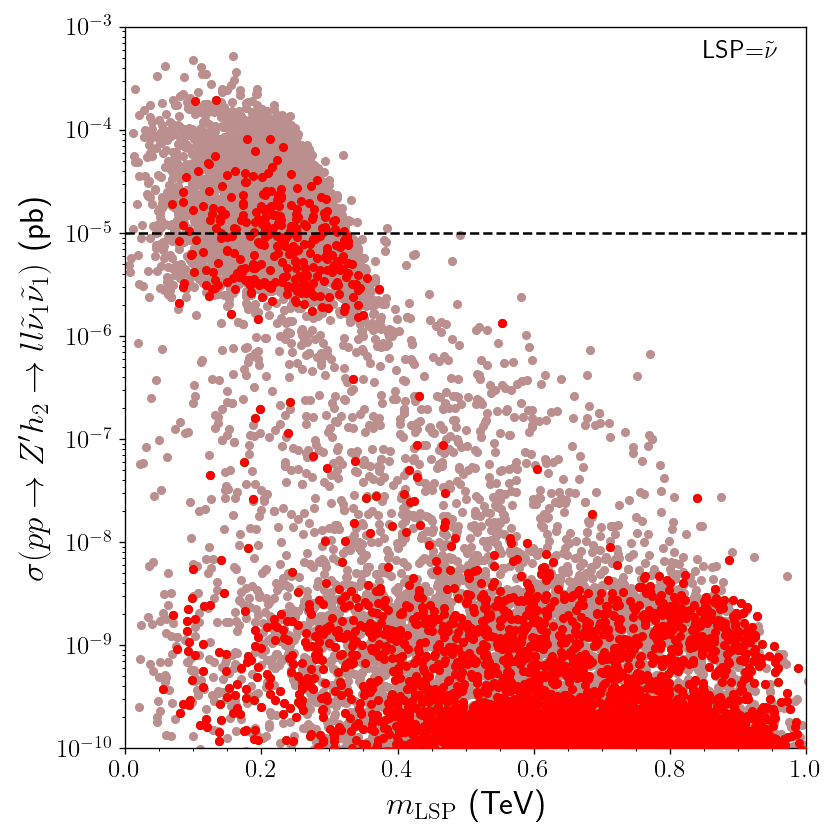}}
\subfigure{\includegraphics[width=6cm,height=5cm]{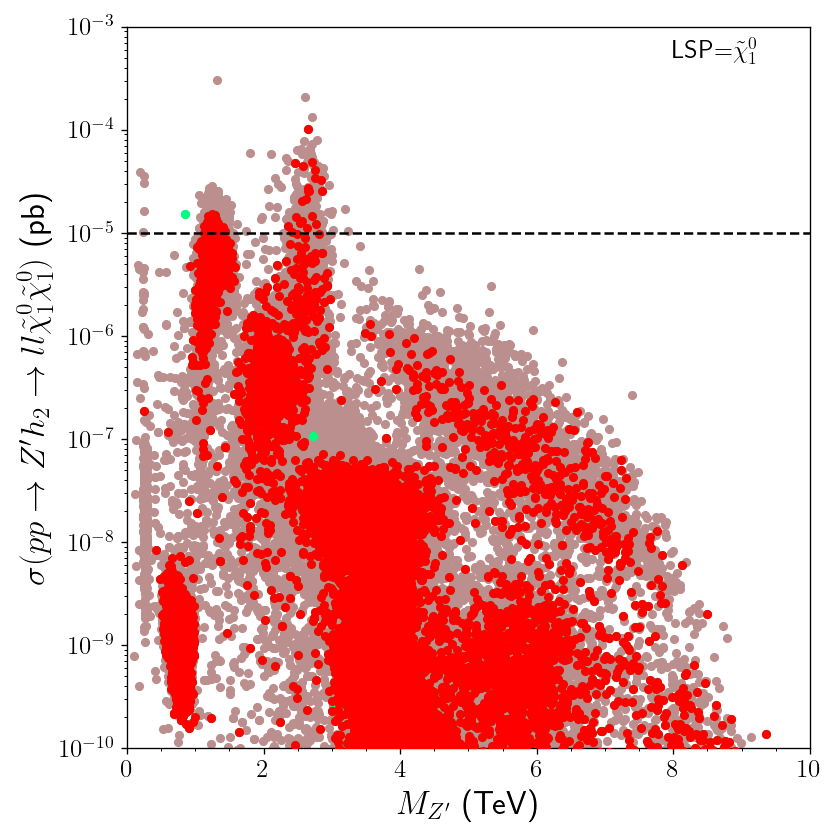}}~~~
\subfigure{\includegraphics[width=6cm,height=5cm]{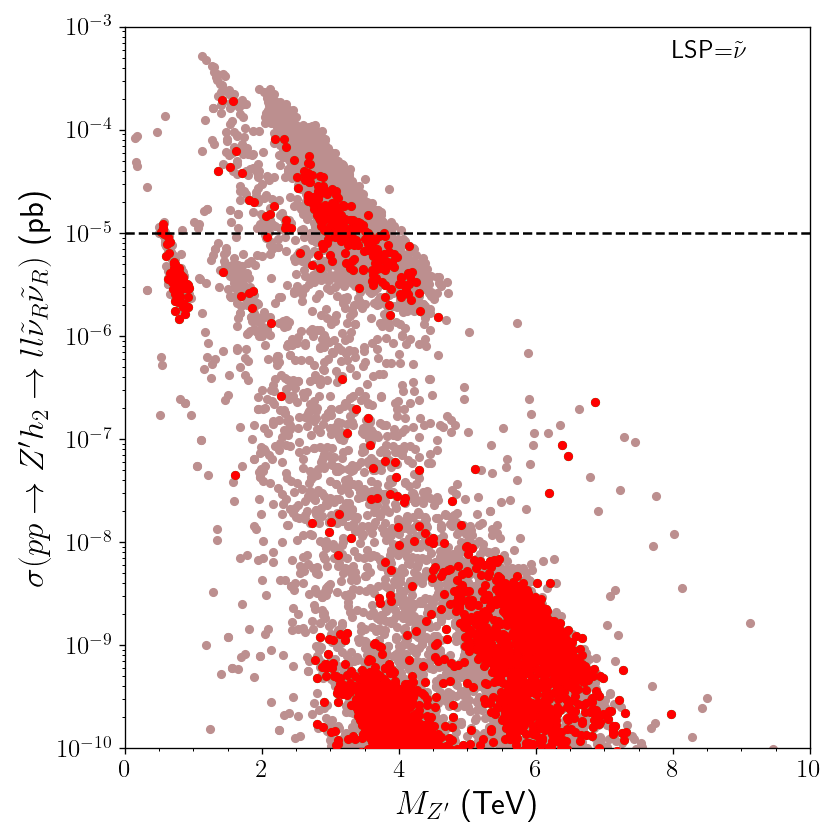}}
\caption{The cross-sections of the signal processes in correlation with the LSP (top) and $Z^{\prime}$ mass (bottom). 
The left panels correspond to scenarios with a neutralino LSP while the right panels show results for a sneutrino LSP.}
\label{fig:signalLSPZP}
\end{figure}

In this section, we present the results for the signal cross-sections as a function of  the masses of the LSPs as well as that of the \( Z^{\prime} \) boson. Figure~\ref{fig:signalLSPZP} shows the signal cross-sections for leptonic final states as a function of the LSP mass (top) and \(Z^{\prime}\) mass (bottom). All points are consistent with  EW Symmetry Breaking (EWSB) and the neutral LSP condition. The shaded points are excluded at least one of the constraints listed in Eq.~(\ref{eq:constraints}), while the red points are consistent with all the constraints. The horizontal dashed lines indicates the solutions for which the cross-section of the signal process is realized at $10^{-5}$ pb or more. These lines are used to separate the solutions which can be probed in current and/or near future collider experiments. The results are presented for both a neutralino LSP (left panels)  and sneutrino  (right panels) LSP.

Despite the presence of multiple Higgs bosons in the spectrum, the signal cross-section can be sizable when the second lightest CP-even Higgs boson (\( h_2 \)) is predominantly composed of singlet fields responsible for \( U(1)_{B-L} \) breaking. As seen in the top panels, cross-sections exceeding \( 10^{-5}~\mathrm{pb} \) are achieved for neutralino LSP masses below approximately 200~GeV. Although less common, sneutrino LSPs can reach masses up to about 400~GeV within a similar cross-section range. For neutralino LSPs, significant cross-sections are found when \( M_{Z^{\prime}} \lesssim 2~\mathrm{TeV} \), although viable rates  also exist for \( Z^{\prime} \) masses up to 3.5~TeV and beyond. In contrast, sizable cross-sections for sneutrino LSPs arise only when \( M_{Z^{\prime}} \lesssim 4~\mathrm{TeV} \).

As intimated, after identifying the solutions with relatively large signal cross-sections, we select two BPs for detailed detector-level analyses, as listed in Table~\ref{tab1}. These BPs are chosen to yield similar values for key observables such as the LSP mass, as previously explained. Furthermore, in view of the forthcoming signal-to-background analysis, we also require the two signal cross-sections, computed at Leading Order (LO), to be similar (within a factor of 2 or so). The only distinguishing feature between these two BPs is thus the nature of the LSP: one corresponds to a neutralino LSP while the other features a sneutrino LSP.

\begin{table}[h!]
\centering
\caption{BPs for the detector analyses, selected to be consistent with all constraints discussed in the text. The masses are given in GeV and the cross-section is expressed in pb.}

\vspace{0.3cm}
\setstretch{2}
\scalebox{0.9}{
\begin{tabular}{|c|c|c|c|} \hline
&  BP1 (LSP \( \tilde{\nu}_{R} \))  & BP2 (LSP \( \tilde{\chi}_{1}^{0} \)) \\ \hline
\( g_{BL} \) & 0.29 & 0.29\\
\( g^{\prime} \) & 0.35 & $-1.48$ \\ \hline
\( m_{h_{\rm SM}} \) & 123.3 & 126.7\\
\( m_{h^{\prime}} \) & 523 & 498.4 \\
\( M_{Z^{\prime}} \) & 2358 & 2361 \\
\hline
\( m_{\tilde{\nu_{R}}} \) & {196.8} & 1592 \\
\( m_{\tilde{\chi}^{0}_{1}} \) & 469.4 & {206.4} \\
\hline
\( \sigma(pp \rightarrow Z^{\prime} h^{\prime} \rightarrow l \bar{l} + \cancel{E}_T) \) & $2.18\times 10^{-5}$ & $4.81\times 10^{-5}$\\ \hline
\end{tabular}}
\label{tab1}
\end{table}

\section{Simulation and Analysis Strategy}
\label{Sec:SimAnalysis}
In the previous sections, the basic structure of the BLSSM-IS framework, the $Z'$, $h'$ and LSP properties as well as the parameter space obtained under phenomenological constraints were discussed in detail. As a result of  scans made, two representative BPs were selected. These points give a meaningful mono-$Z^\prime$ signal and are consistent with the main experimental constraints. As explained already, they differ from each other in terms of the LSP candidate appearing in the invisible final state. In the BP1 scenario, the LSP is the right-handed sneutrino while BP2 adopts the neutralino as LSP.

As discussed in Section~\ref{sec:intro}, in mono-$Z^\prime$ production, the $Z^\prime$ boson  decays into two charged leptons while the $h^\prime$ boson  decays into two stable LSP. In this section, these processes are studied at detector level in a hadron collider  for both BPs and their distinguishability against the dominant SM backgrounds is investigated.

The relevant signal processes are given in Eq.~(\ref{SimAnSt:eq1}). 
\begin{align}
\text{BP1}:\qquad  &p\; p \rightarrow Z^{\prime}  \; h^\prime,\; (Z^{\prime}  \rightarrow l^+ l^-),(h^\prime \rightarrow \tilde{\nu_1}\tilde{\nu_1}),\qquad (l^\pm: e^\pm,\mu^\pm) \nonumber \\
\text{BP2}:\qquad  &p\; p \rightarrow Z^{\prime}  \; h^\prime,\; (Z^{\prime}  \rightarrow l^+ l^-),(h^\prime \rightarrow \tilde{\chi}_1^0\tilde{\chi}_1^0)
\label{SimAnSt:eq1}
\end{align}
The dominant SM background processes are given in Table.~\ref{tab:sb_processes}.

The signal and all SM background processes were generated at the parton level using MadGraph5\_aMC@NLO (v3.6.3)  \cite{Alwall:2014hca}. The 5-Flavour Scheme (5FS) and NNPDF23LO1 Parton Distribution Functions (PDFs) \cite{NNPDF:2017mvq}
(with default renormalization/factorization scale choice) were employed. The generated events were then processed through Pythia 8.3 \cite{Sjostrand:2014zea} for showering, hadronization and decay of unstable particles. Detector simulation was performed using the Delphes 3.5.0 software \cite{deFavereau:2013fsa}. The CMS detector card delphes\_card\_CMS.tcl is the one used. In this setup, electrons are reconstructed with $p_T^e > 10~\mathrm{GeV}$ and $|\eta_e| < 2.5$ while muons satisfy $p_T^\mu > 10~\mathrm{GeV}$ and $|\eta_\mu| < 2.4$. Jets are reconstructed with the anti-$k_T$ algorithm with $R = 0.5$ and $p_T^j > 20~\mathrm{GeV}$. Standard lepton isolation criteria, as implemented in the CMS Delphes card, are applied to leptons. More specifically, a cone size of $\Delta R = 0.5$ is used, and only surrounding energy-flow objects with $p_T > 0.5~\mathrm{GeV}$ are included in the isolation. The corresponding isolation requirements are set to 0.12 for electrons and 0.25 for muons. We use $\sqrt s=14$ TeV for the LHC energy (and eventually a luminosity of ${\cal L}=$ 139 fb$^{-1}$ (i.e., the current Run 3 dataset) and 3 ab$^{-1}$ (i.e., the High-Luminosity LHC (HL-LHC) dataset \cite{Gianotti:2002xx}).

\begin{table}[h!]
  \centering

  \begin{tabularx}{\textwidth}{>{\raggedright\arraybackslash}Xcc}
    \toprule
    Background Process & $\sigma_{\text{LO}}$ [14 TeV] (pb) & K-Factor [14 TeV] \\
    \midrule

    \makecell[l]{$pp \rightarrow t \bar{t},\ (t \rightarrow W^+ b,\ (W^+ \rightarrow l^+ \nu_l)),$ \\
    $(\bar{t} \rightarrow W^- \bar{b},\ (W^- \rightarrow l^- \bar{\nu}_l))$}
    & $29.68$ & $1.72$ \\

    \makecell[l]{$pp \rightarrow t W^\pm,\ (t \rightarrow W^+ b,\ (W^+ \rightarrow l^+ \nu_l)),$ \\
    $(W^\pm \rightarrow l^\pm \nu_l)$}
    & $3.227$ & $1.415$ \\

    \makecell[l]{$pp \rightarrow t \bar{t} V\ (V = Z, W^\pm),\ (Z \rightarrow \nu_l \bar{\nu}_l),$ \\
    $(t \rightarrow W^+ b,\ (W^+ \rightarrow l^+ \nu_l)),$ \\
    $(\bar{t} \rightarrow W^- \bar{b},\ (W^- \rightarrow l^- \bar{\nu}_l))$}
    & $0.01154$ & $1.48$ \\

    \makecell[l]{$pp \rightarrow VV\ (V = Z, W^\pm),\ (W^\pm \rightarrow l^\pm \nu_l),$ \\
    $(Z \rightarrow \nu_l \bar{\nu}_l)$}
    & $1.904$ & $1.599$ \\

%    \makecell[l]{$pp \rightarrow Z/\gamma^*$, \\
%    Final state $(2\,l^\pm + 4\,\nu_l)$}
%    & $9.178 \times 10^{-7}$ & $1.498$ \\

    \bottomrule
  \end{tabularx}
  \caption{LO cross-sections and NLO K-factors for backgrounds.}
    \label{tab:sb_processes}
\end{table}

%\begin{table}[h!]
%  \centering
%  \caption{LO cross-sections and NLO K-factors for backgrounds.}
%  \label{tab:sb_processes}
%  \resizebox{\textwidth}{!}{%
%  \begin{tabular}{lcc}
%    \toprule
% Background Process & $\sigma_{\text{LO}}$ [14 TeV] (pb) & K-Factor [14 TeV] \\
%    \midrule
%  $pp \rightarrow t \bar{t}, (t \rightarrow W^+ b, (W^+ \rightarrow l^+ \nu_l)),(\bar{t}  \rightarrow W^- \bar{b}, (W^- \rightarrow l^- \bar{\nu_l})) $ & $29.68$ & $1.72$  \\
%   $pp \rightarrow t W^\pm, (t \rightarrow W^+ b, (W^+ \rightarrow l^+ \nu_l)),(W^\pm \rightarrow l^\pm \nu_l)$ & $3.227$& $1.415 $\\
%   $pp \rightarrow t \bar{t} V (V = Z, W^\pm), (Z\rightarrow \nu_l \bar{\nu_l}),(t \rightarrow W^+ b, (W^+ \rightarrow l^+ \nu_l)),(\bar{t}  \rightarrow W^- \bar{b}, (W^- \rightarrow l^- \bar{\nu_l}))$ & $ 0.01154$ & $1.48$  \\
%   $pp \rightarrow VV (V = Z, W^\pm), (W^\pm \rightarrow l^\pm \nu_l),(Z\rightarrow \nu_l \bar{\nu_l}) $ &$1.904$ & $1.599$\\
%   $pp \rightarrow Z/\gamma^*$, Final state ($2 l^\pm + 4 \nu_l$) & $9.178 \times 10^{-7}$& $1.498$ \\
%    \bottomrule
%  \end{tabular}
%  }
%\end{table}

The computational chain used for event generation, parton showering, hadronization and detector simulation is then run. In  signal events, the final state contains two oppositely charged leptons of same flavor together with Missing Transverse Energy (MET or $\rm {E}_T^{\rm Miss}$), here, due to the pair of DM particles. For this reason, the di-boson processes $W^+W^-$, $W^\pm Z$ and $ZZ$ form irreducible backgrounds, since they can produce such two leptons and neutrinos in the final state\footnote{Processes coming from $Z/\gamma^*$ can also give a background very close to the signal topology when they produce a lepton pair together with neutrinos, yet these were found to be negligible}. In addition, processes involving top (anti)quarks, such as $t\bar{t}$, $tW^\pm$ and $t\bar{t}V$, are also important background sources. Through leptonic decay chains, these processes produce isolated leptons and neutrinos, and therefore also lead to missing energy. Background samples were normalized using NLO K-factors from Refs. \cite{Kidonakis:2023juy,Kim:2024ppt}, while the signal was kept at LO were used. For the signal, no K-factor was applied, and the cross section was kept at LO.

%Using K-factors for the background processes is important in order to obtain a more realistic normalization in the analysis. In this work, both the signal and background samples were generated at LO using MadGraph. However, because of QCD corrections, electroweak contributions, and higher-order virtual and real emissions, the physical cross sections can differ noticeably from the LO predictions. For this reason, K-factors taken from (Next to Leading Order) NLO results in the literature were also applied, so that the normalization of the LO samples generated with MadGraph becomes more realistic. For the LHC at 14~TeV, the values $K_{t\bar t}=1.72$~\cite{Kidonakis:2023juy}, $K_{tW}=1.415$, $K_{t\bar tV}\simeq 1.48$ and $K_{VV}=1.599$
%, and $K_{Z/\gamma^*}=1.498$~\cite{Kim:2024ppt} 

\subsection{Event Selection}

For the BP1 and BP2 signal scenarios, the lepton pair is produced by the decay of a heavy resonance. On the one hand, the dilepton system typically represents the visible component of a very hard event topology. On the other hand, the missing transverse energy, $\rm E_T^{\rm Miss}$, is produced by invisible particles. The sum of three momenta of the invisible particles is not necessarily large. In fact, the transverse momentum vectors of the two invisible particles may cancel each other. This means that the $\rm E_T^{\rm Miss}$ is small. This is particularly relevant since, even if the $p_T$ of the dilepton system is large, the $\rm E_T^{\rm Miss}$ may not increase accordingly. This makes it  harder to obtain a large separation between the signal and the background. In the two BPs under consideration, the visible component of the signal at the detector level is the same:

\begin{equation}
pp \to Z^\prime h^\prime, \qquad Z^\prime \to l^+l^-, \nonumber
\end{equation}
with the $h^\prime$ Higgs boson decaying invisibly into the two DM states
\begin{equation}
h^\prime \to \tilde{\nu}_1 \tilde{\nu}_1 \quad \text{(BP1)}, 
\qquad
h^\prime \to \tilde{\chi}_1^0 \tilde{\chi}_1^0 \quad \text{(BP2)}. \nonumber
\end{equation}

Although the invisible final-state particles are physically different here, this  does not lead to a directly distinguishable effect at the detector level in the present analysis. Any potentially distinguishing information coming from the spin structure is not efficiently transferred to the visible final state. The main reason is that these invisible particles are produced through the decay of an intermediate scalar particle, namely $h^\prime$. Since $h^\prime$ is a scalar, its decay does not generate a preferred spin orientation, hence, a strong angular pattern. From the experimental side, the available information is just from the dilepton system coming from the decay of the $Z^\prime$ and the vector sum of the transverse momenta of the invisible particles. 

The remaining differences between BP1 and BP2 are potentially due to kinematic effects associated with the different masses involved for both the $Z'$ and DM state, which we have selected being similar for both so as to minimize such effects,    whereas production cross-sections and BRs, while different for the two BPs, they just induce a different signal rate overall. The purpose of our MC analysis it to establish a selection which is insensitive to the nature of DM, thereby establishing a search channel offering scope for its spin-independent extraction. 

\begin{figure}[t]
\centering
\subfigure{\includegraphics[width=6cm,height=5cm]{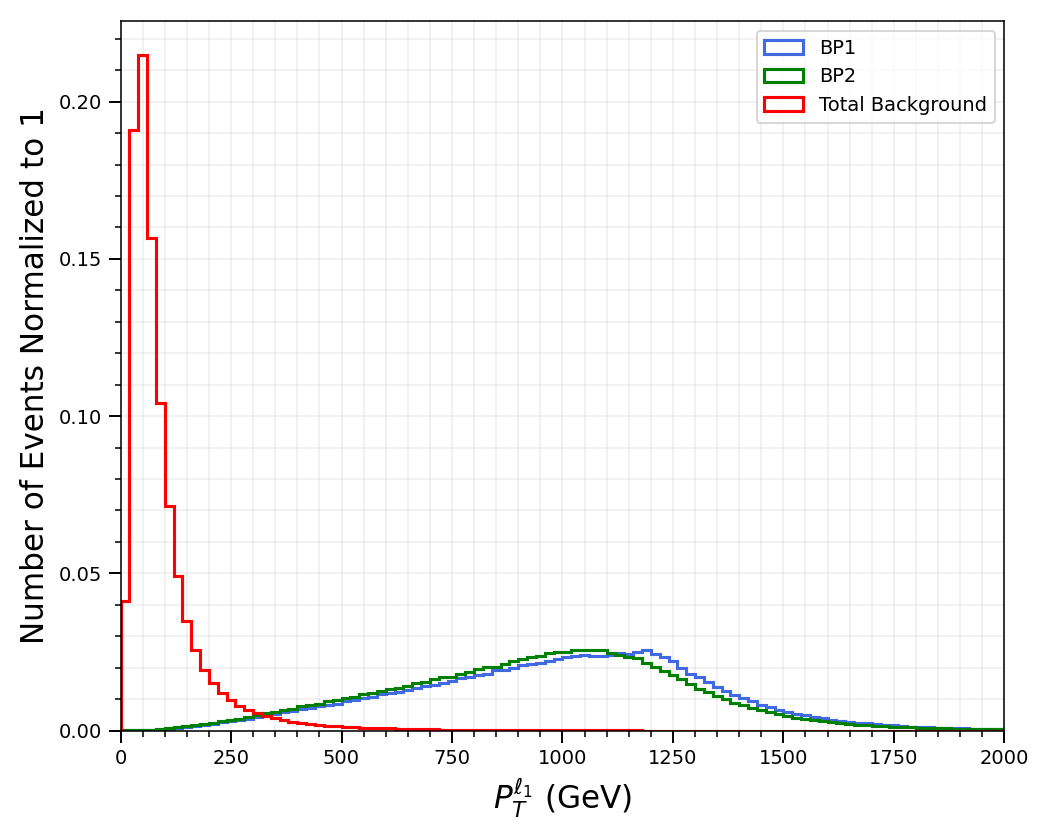}}
\subfigure{\includegraphics[width=6cm,height=5cm]{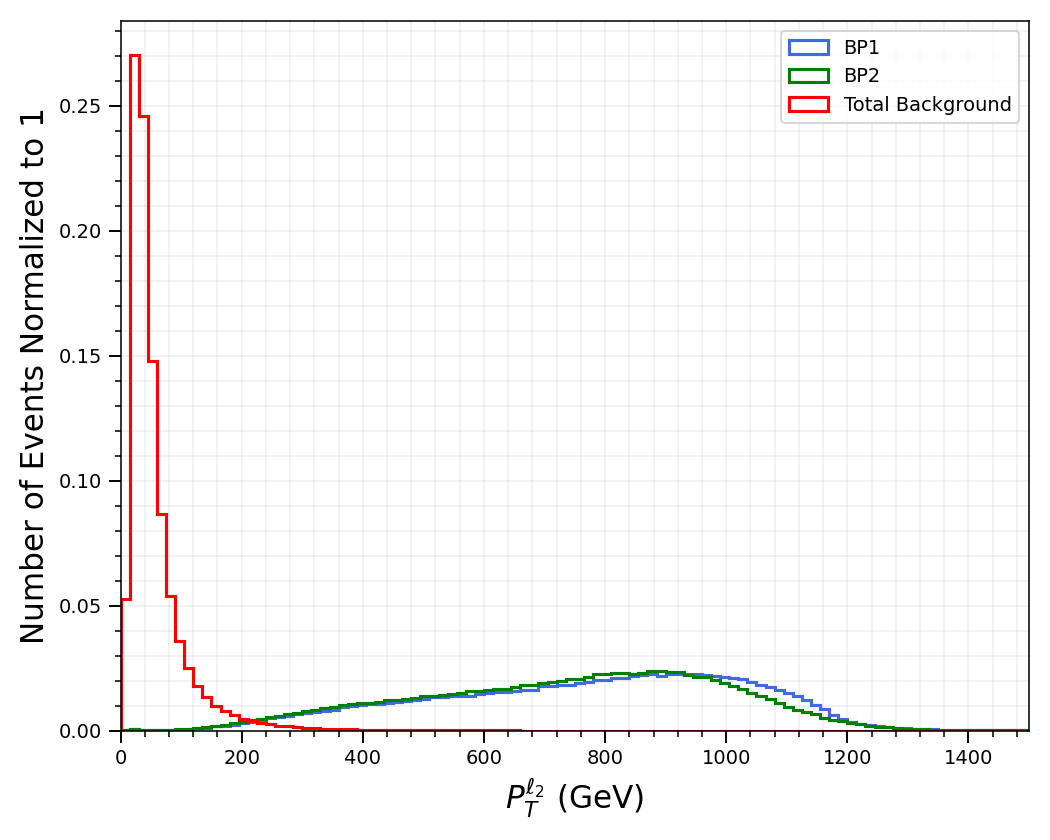}}
\subfigure{\includegraphics[width=6cm,height=5cm]{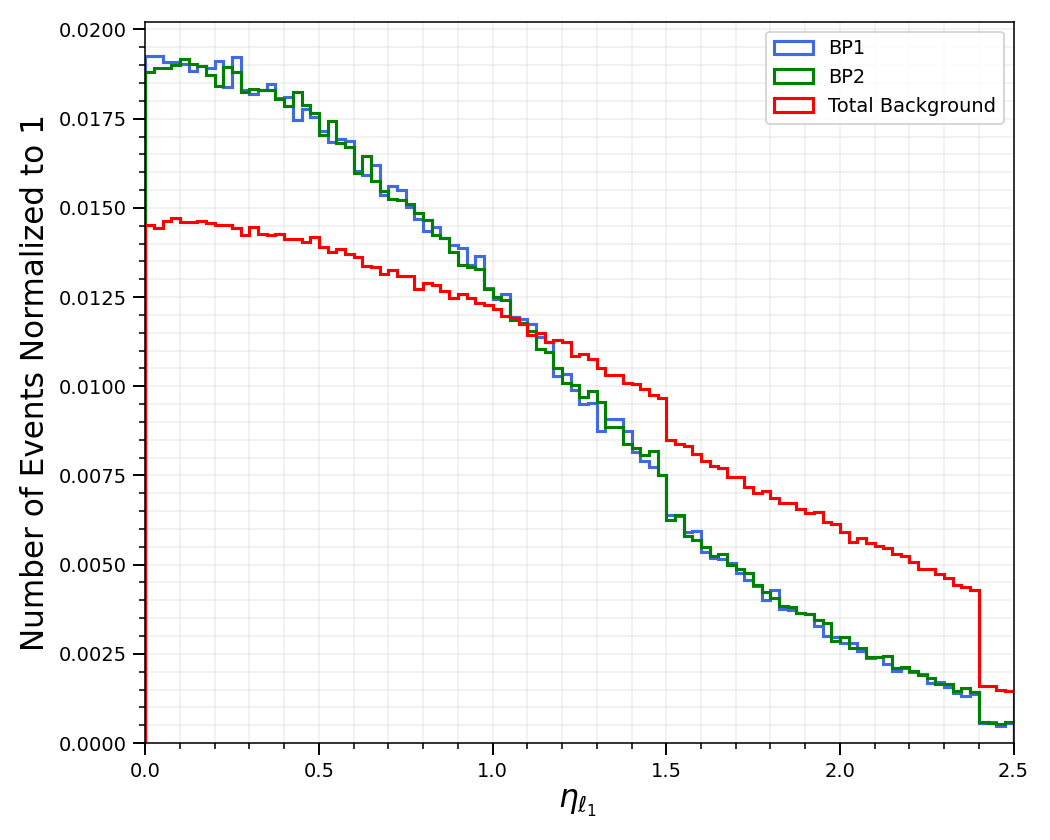}}
\subfigure{\includegraphics[width=6cm,height=5cm]{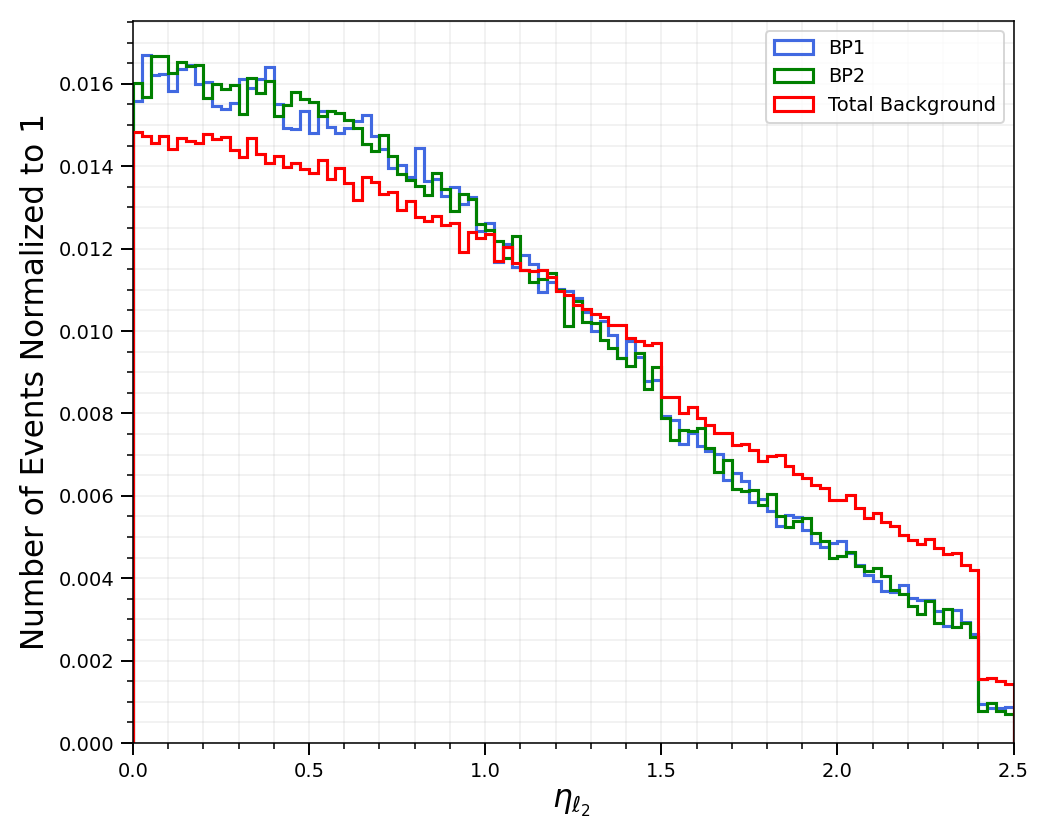}}
\caption{Normalized kinematic distributions in the transverse momentum (top) and pseudorapidity (bottom) of the leading (left) and subleading (right) leptons for BP1, BP2 and  total SM background.}
\label{fig:lep_kinematics}
\end{figure}

Let us come to the backgrounds now. The most powerful discriminant between these and either of the signals is the invariant mass of the two leptons. Before turning to the (final) invariant mass selection, though, it is useful to examine the normalized distributions of the main kinematic observables characterizing the visible leptonic system, so as to exploit them in our selection. Figure \ref{fig:lep_kinematics} shows the transverse-momentum and pseudorapidity distributions of the leading ($l_1$) and subleading ($l_2$) leptons for BP1, BP2 and the total SM background.

As seen in Figure~\ref{fig:lep_kinematics} (top-left) and Figure~\ref{fig:lep_kinematics} (top-right), the signal distributions for both the leading  and subleading leptons in the signals are much harder than those of the total background. This behavior is directly related to the large mass of the parent resonance, $m_{Z'} \simeq 2.36$~TeV, so that the leptons produced in the $Z' \to l^+l^-$ decay inherit transverse momenta at a scale ranging from several hundreds of GeV up to the TeV range. By contrast, the background leptons are mostly generated through $W^\pm$, $Z$, and top (anti)quark decay chains, which are characterized by substantially lower energy scales. The similarity between BP1 and BP2 is expected, since the visible dilepton system is produced in the same way in both scenarios, while the difference between the two BPs is mainly associated with the invisible sector (as already pointed out). 
The pseudorapidity distributions in Figure~\ref{fig:lep_kinematics} (bottom-left) and Figure~\ref{fig:lep_kinematics}(bottom-right) show that the signal leptons are produced somewhat more centrally than in the total SM background. This is  consistent with the decay of a heavy $s$-channel resonance into a dilepton final state whereas the topologies characterizing the background processes are very varied.
Although no additional cut is imposed directly on these observables beyond the aforementioned detector acceptances, they confirm that significant differences between signal and background events that can be used for interpretation purposes (of a possible excess).

As for the preselection criteria to be used in our analysis, these were chosen to suppress the backgroundz at an early stage while preserving the signal regions as much as possible. To start with, the two leptons are required to have opposite same flavor and opposite electric charge. This condition helps reduce the combinatorics of the backgrounds emerging from diboson states in the decay chains. A $b$-jet veto is also applied at the preselection level, as in top (anti)quark induced backgrounds (i.e., $t\bar{t}$, $tW^\pm$ and $t\bar{t}V$), detectable $b$-jets are very often present in the final state. In contrast, no obvious  $b$-jet signature is expected in the signal process considered here. Therefore, applying a $b$-jet veto is highly effective. Finally, a requirement of $E_T^{\rm Miss} \geq 20~\mathrm{GeV}$ is imposed. Since the $h^\prime$ boson decays invisibly in the signal process into rather massive objects (pair of DM candidates), some sizable amount of MET is naturally expected in the signals. For this reason, even a modest threshold on $E_T^{\rm Miss}$ is physically well motivated. This requirement also helps reduce artificial low $E_T^{\rm Miss}$ contributions arising from detector fluctuations or poorly measured events. 
%The overall purpose of these preselection criteria is to reduce, at an early stage, backgrounds that are mainly top (anti)quark induced or characterized by low missing energy, while keeping the basic structure of the signal intact. 

%The preselection variables that most directly motivate the early background rejection are the missing transverse energy and the number of $b$-tagged jets. Their normalized distributions are shown in Figure~\ref{fig:preselection_observables}. These distributions clarify why a modest $\rm {E}_T^{\rm Miss}$ requirement together with a $b$-jet veto is already effective before the final invariant-mass cut is imposed.

\begin{figure}[t]
\centering
\subfigure{\includegraphics[width=6cm,height=5cm]{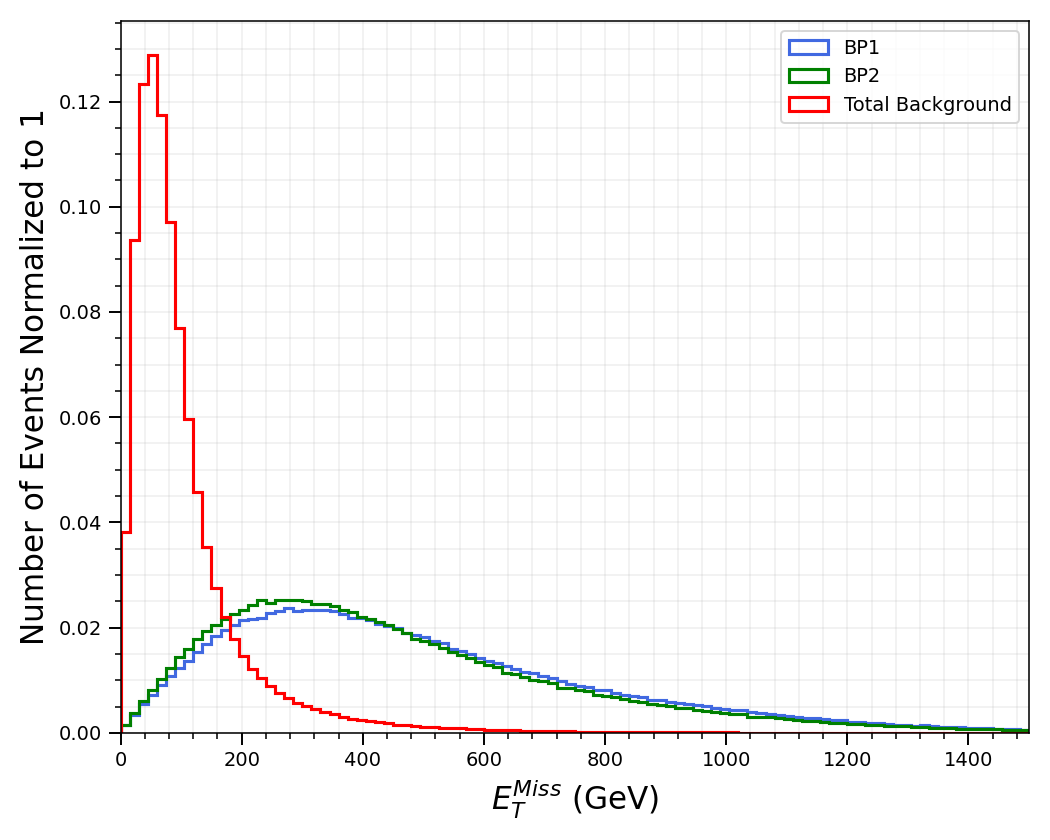}}
\subfigure{\includegraphics[width=6cm,height=5cm]{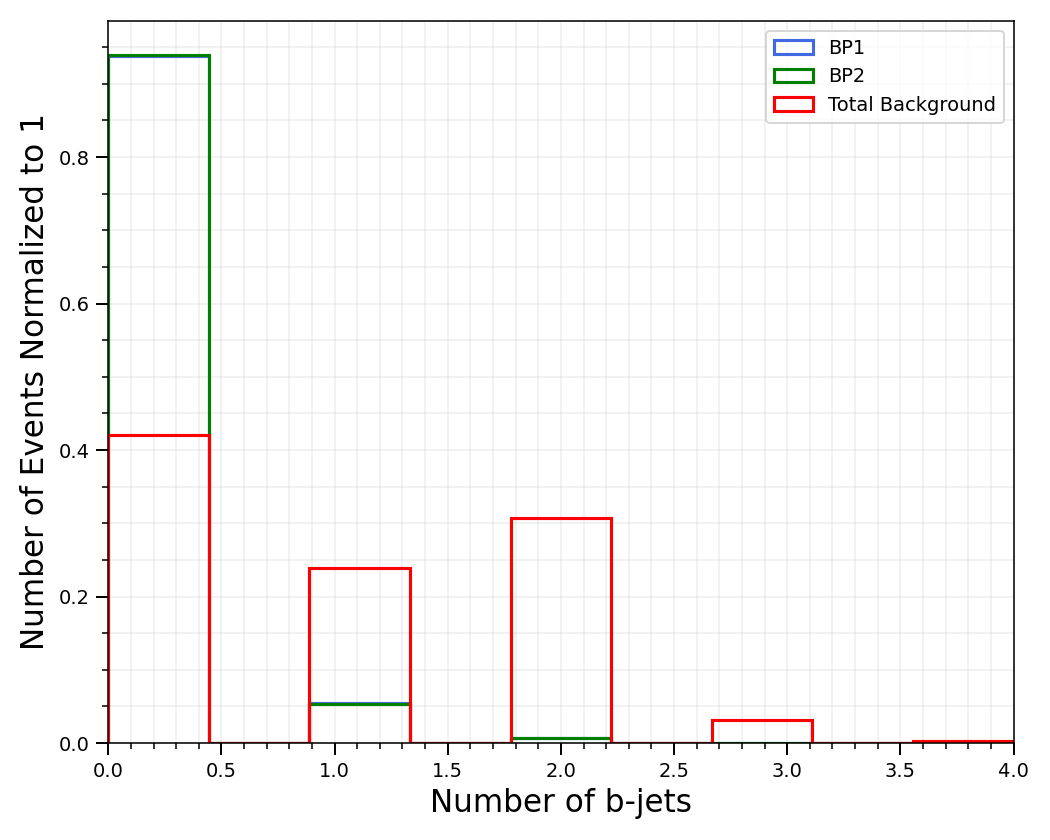}}
\caption{Normalized kinematic distributions in the MET (left) and $b$-jet multiplicity for BP1, BP2 and  total SM background.}
\label{fig:preselection_observables}
\end{figure}

Normalized distributions of the discussed preselection variables are presented in Figure~\ref{fig:preselection_observables}.  Firstly, as shown in Figure~\ref{fig:preselection_observables} (left), the signal BPs clearly differ from the total background in terms of the $\rm {E}_T^{\rm Miss}$ distribution, as intimated. That is, the latter distribution peaks in the low $\rm {E}_T^{\rm Miss}$ domain while the former spectrum extends up to relatively high values of $\rm {E}_T^{\rm Miss}$. %This follows from the fact that there are two invisible LSPs among the decay products of the $h^\prime$ boson. 
Yet, the signal is not excessively localized towards $\rm {E}_T^{\rm Miss} \gg 100$~GeV since the transverse momenta of the two invisible particles may compensate each other (as previously mentioned). This makes a cut as $\rm {E}_T^{\rm Miss} > 20$~GeV quite motivated and, indeed, it removes background events significantly while retaining a substantial fraction of signal ones.
Secondly,  Figure~\ref{fig:preselection_observables} (right) demonstrates the physical motivation for applying the $b$-jet veto to the data. Namely, in the signal benchmark sample, the number of events is predominantly contained in the $N_b = 0$ bin, which is obviously related to the absence of intrinsic $b$-quarks in this sample. The remaining part of the signal distribution in bins $N_b \geq 1$ can appear due to QCD radiation, heavy flavor production and shower mistaggings. In contrast, the total background quite substantial in the $N_b = 1, 2$ bins, which is the result of the significant contribution from backgrounds including top (anti)quarks, i.e., the $t\bar{t}$, $tW^\pm$ and $t\bar{t}V$ channels.

\begin{figure}[t]
\centering
\includegraphics[width=6cm,height=5cm]{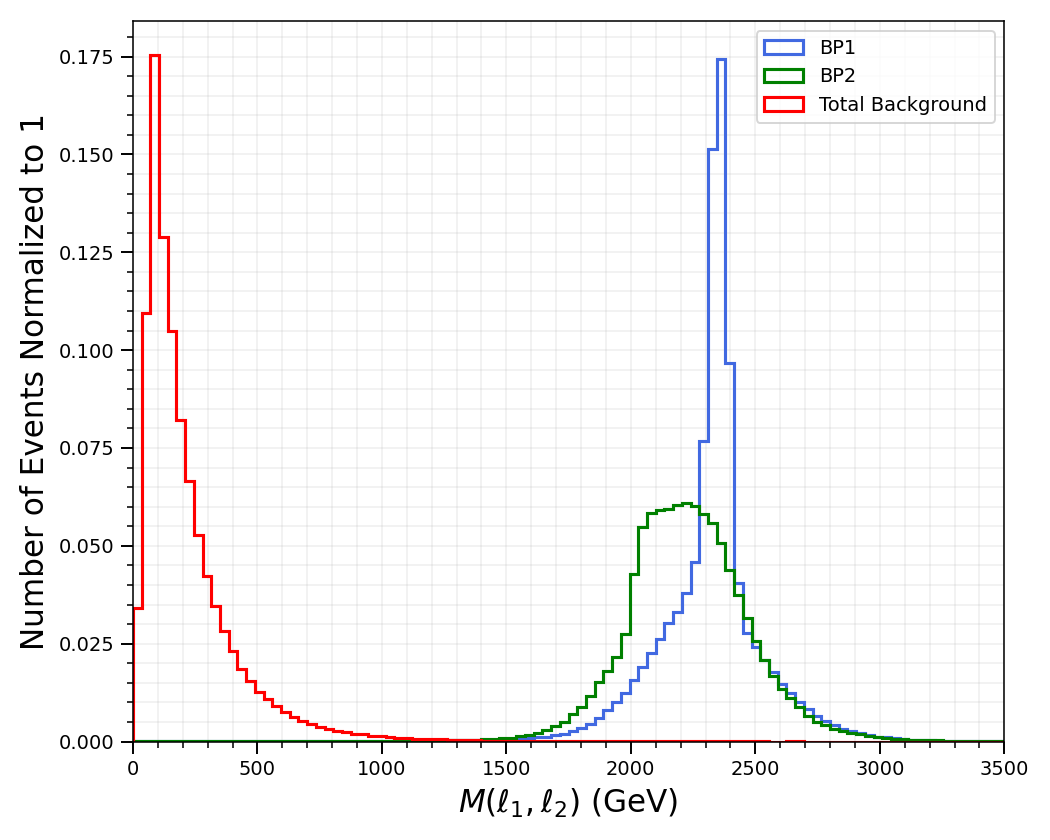}
\caption{Normalized kinematic distribution in the dilepton invariant mass distribution for BP1, BP2 and total SM background.}
\label{fig:mll_dist}
\end{figure}

A specific veto on the dilepton invariant mass around $M_Z$ is not enforced due to the fact that the requirement of $M({l_1,l_2})>1250~\mathrm{GeV}$ that we are finally enforcing  automatically removes any possibility of being within the $Z$ on-shell phase space region. No particular veto on the mass of the top (anti)quark is considered either since the background contributions arising from such processes are already heavily suppressed through the application of the $b$-jet veto and the dilepton mass cut. As for $W^\pm$ related backgrounds, although no explicit $W^\pm$ mass inspired veto is implemented, they are efficiently reduced by the same high $M({l_1,l_2})$ requirement, which ultimately provides the dominant discrimination power in the present analysis.
In fact, Figure \ref{fig:mll_dist} displays the dilepton invariant mass distribution 
(i.e., $(M({l_1,l_2}) \equiv \sqrt{(p_{l_1}+p_{l_2})^2})$)
for BP1, BP2 and  total SM background. We notice that the latter are clustered at low values of $M({l_1,l_2})$ while  the signal events are concentrated at the higher end of the invariant mass, roughly between 2.0 and 2.5 TeV. This makes the aforementioned cut $M({l_1,l_2})>1250$  GeV particularly useful because this requirement keeps the maximum signal and reduces the background significantly. % This threshold is really important. It should not be too close to the resonance peak. If we set the cut too high we will lose a lot of signal especially when we are dealing with the resonance peak. We have a smaller number of events or a narrower distribution. The resonance peak is important here. If we set the cut too low we will not be able to suppress the background. So the resonance peak and the threshold are important. The value of $1250~\mathrm{GeV}$ is a balance, between the resonance peak and the signal efficiency and the background rejection and the resonance peak.

The shape of the distribution is not determined by the resonance mass alone, as it also reflects the characteristics of the propagator itself, which has a Breit-Wigner form. The behavior of the squared amplitude near the resonance can be approximated as
\begin{equation}
    |D(\hat{s})|^2 \propto 
\frac{1}{(\hat{s}-m_{Z^\prime}^2)^2 + m_{Z^\prime}^2 \Gamma_{Z^\prime}^2}\,. \nonumber
\end{equation}
It can be noted that the invariant dilepton mass distribution for BP2 has a larger width, $\Gamma = 349.07$ GeV, compared to that of BP1, $\Gamma = 52.66$ GeV. Therefore, the dilepton invariant mass distribution for BP2 has a larger spread compared to that of BP1, so that the shape differences observed in the invariant dilepton mass distribution in the last figure can be attributed to the differences in the width of the resonance peaks. Such an effect is another reason why we have decided to place a minimal value on $M({l_1,l_2})$ rather than requiring a window around the $Z'$ mass, so that the latter needs not be know to
perform the advocated signal-to-background selection. (In fact, it should further be noted that no assumption is made on the $h'$ mass either.)

\subsection{Outcomes of the Detector Level Analysis}

\begin{table}[t!]
  \centering
  \renewcommand{\arraystretch}{1.15}

  \begin{minipage}{\textwidth}
    \centering
    \setlength{\tabcolsep}{4pt}
    \begin{tabular}{lcccccc}
      \toprule
      Cuts &
      \makecell{Signal\\(BP1)} &
      \makecell{Signal\\(BP2)} &
      $t\bar{t}$ &
      $t\bar tV$ &
      $tW^\pm$ &
      $VV$   
      %& $Z\gamma$ 
      \\
      \midrule
      Before Cuts &
      3.033 & 6.680 &
      $7.096 \times 10^{6}$ & 2374.009 & $6.347 \times 10^{5}$ &
      $4.232 \times 10^{5}$
      %& 0.191 
      \\
      Preselection &
      2.577 & 3.646 &
      $4.643 \times 10^{4}$ & 13.703 & $2.034 \times 10^{4}$ &
      $8.763 \times 10^{4}$ 
      %& 0.142 
      \\
      $M(l_1,l_2) > 1250~\mathrm{GeV}$ &
      1.378 & 3.627 &
      0 & 0.00833 & 0 & 11.285 
      %& 0.00052 
      \\
      \bottomrule
    \end{tabular}
  \end{minipage}

  \vspace{0.6cm}

  \begin{minipage}{\textwidth}
    \centering
    \setlength{\tabcolsep}{6pt}
    \begin{tabular}{lccccc}
      \toprule
      Cuts &
      \makecell{Total\\background} &
      \makecell{$\epsilon_S$\\(BP1)} &
      \makecell{$\epsilon_S$\\(BP2)} &
      \makecell{$\mathcal{Z}$\\(BP1)} &
      \makecell{$\mathcal{Z}$\\(BP2)} \\
      \midrule
      Before Cuts &
      $8.156 \times 10^{6}$ &
      1.000 & 1.000 &
      {0.00106} & {0.00234} \\
      Preselection &
      $1.544 \times 10^{5}$ &
      0.850 & 0.546 &
      0.00656 & 0.00928 \\
      $M(l_1,l_2) > 1250~\mathrm{GeV}$ &
      11.294 &
      0.454 & 0.543 &
      {0.387} & {0.939} \\
      \bottomrule
    \end{tabular}
  \end{minipage}

  \caption{Cutflow at $\sqrt{s}=14$ TeV with $\mathcal{L}=139 \;\mathrm{pb}^{-1}$. The number of events for the signals and backgrounds are given. In addition, $\epsilon$ represents the cut efficiency for BP1 and BP2 and $\mathcal{Z}$ represents their significance.}
  \label{tab:cutflow}
\end{table}

%\begin{table}[h!]
%  \centering
%  \resizebox{\textwidth}{!}{%
%  \begin{tabular}{lcccccccccccc}
%    \toprule
%    Cuts &
%    Signal (BP2) & Signal (BP1) &
%    $t\bar{t}$ & $t\bar tV$ & $tW$ & $VV$ & $Z\gamma$ &
%    Total background &
%    $\epsilon_S^{\mathrm{BP2}}$ & $\epsilon_S^{\mathrm{BP1}}$ &
%    $\mathcal{Z}^{\mathrm{BP2}}$ & $\mathcal{Z}^{\mathrm{BP1}}$ \\
%    \midrule
%    Before Cuts &
%    6.68034 & 3.03298 &
%    $7.0959 \times 10^{6}$ & 2374.0088 & $6.3470 \times 10^{5}$ & $4.2318 \times 10^{5}$ & 0.191106152 &
%    $8.1562 \times 10^{6}$ &
%    1 & 1 &
%    {0.002339136} & {0.001062005} \\
%    Preselection &
%    3.645806689 & 2.576525 &
%    $4.6431 \times 10^{4}$ & 13.70265422 & $2.0335 \times 10^{4}$ & $8.7630 \times 10^{4}$ & 0.141992508 &
%    $1.5441 \times 10^{5}$ &
%    0.545751667 & 0.849502799 &
%    0.009277952 & 0.006556783 \\
%    $M(l_1,l_2) > 1250~\mathrm{GeV}$ &
%    3.626545042 & 1.377751385 &
%    0 & 0.008332414 & 0 & 11.28493184 & 0.000519172 &
%    11.29378343 &
%    0.542868333 & 0.454256667 &
%    {0.938866588} & {0.387167755} \\
%    \bottomrule
%  \end{tabular}%
%  }
%  \caption{Cutflow at $\sqrt{s}=14$ TeV with $\mathcal{L}=139 \;\mathrm{pb}^{-1}$. The number of events for the signal and background are given. In addition, $\epsilon$ represents the cut efficiency for BP1 and BP2, and $\mathcal{Z}$ represents the significance.}
%  \label{tab:cutflow}
%\end{table}

Table \ref{tab:cutflow} presents the cutflow results for $\sqrt{s}=14$~TeV and $\mathcal{L}=139~\mathrm{fb}^{-1}$. Before applying the cuts, the numbers of signal events for both BP1 and BP2 are quite limited. In contrast, the total background is very large. As a consequence of the preselection, the background is reduced signficantly, though. In particular, processes like $t\bar{t}$ and $tW^\pm$ are heavily suppressed due to the $b$-jet and the request for a minimum $E_T^{\rm Miss}$. However, at this stage, the total SM background is still much larger than the signal. Therefore, the preselection alone is not useful for discovery but helps to give a cleaner sample on which the final cut can work more effectively. The main discriminating step is thus the cut $M(l_1,l_2) > 1250~\mathrm{GeV}$. After it, the $t\bar{t}$ and $tW^\pm$ backgrounds become quite negligible. Only a very small $t\bar tV$ contribution remains, together with the dominant $VV$ background. As a result, the total background is reduced to about $11.29$ events. In contrast, the signal events are considerably preserved. The total signal efficiency remains about $\epsilon_S^{\mathrm{BP2}} \simeq 0.543$ for BP2 and $\epsilon_S^{\mathrm{BP1}} \simeq 0.454$ for BP1. This shows that the invariant mass cut keeps the signal events while providing a very strong suppression of the background. After this final cut, the statistical significance, defined as
\begin{equation}
    \mathcal{Z} = \frac{S}{\sqrt{S+B}},
\end{equation}
where $S$ and $B$ denote the numbers of signal and background events, respectively, is found to be
$\mathcal{Z}^{\mathrm{BP2}} \simeq 0.939$ for BP2 and
$\mathcal{Z}^{\mathrm{BP1}} \simeq 0.387$ for BP1. These values are not sufficient for a discovery in the current LHC experiments. However, they show that the chosen analysis strategy is focusing on the right variables and provides a quite promising starting point, especially for BP2.

\begin{figure}[h!]
  \centering
    \includegraphics[width=0.48\linewidth]{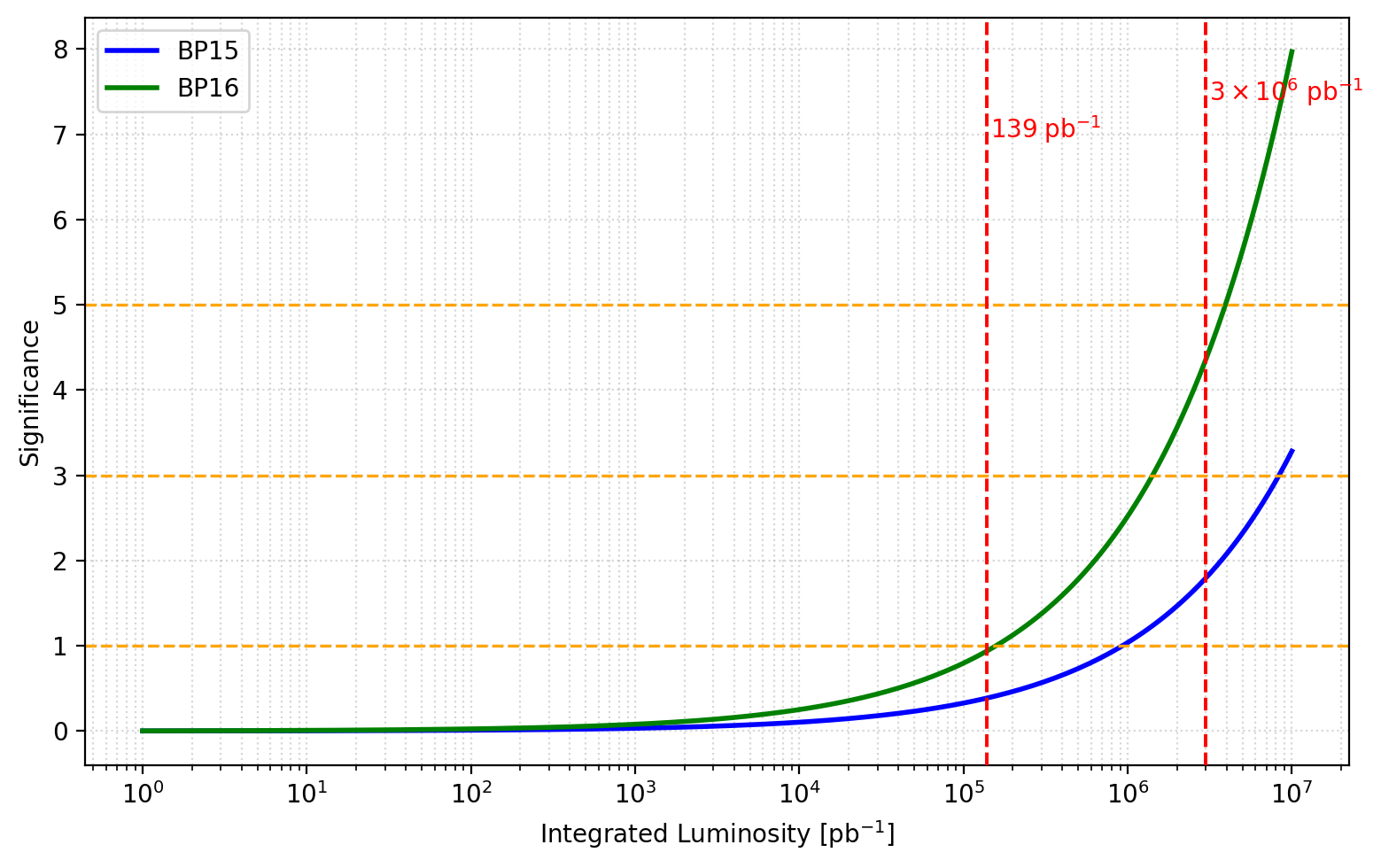}
    \caption{Relationship between luminosity and significance for our signals.}
  \label{fig:SS_vs_L}
\end{figure}

Indeed, Figure \ref{fig:SS_vs_L} indicates that BP2 leads to a larger significance than BP1 across the considered luminosity range. This is mainly due to the larger production rate of BP2 and the correspondingly higher number of selected signal events after the final cuts. Consequently, BP2 approaches a much stronger observability regime at HL-LHC luminosities, whereas BP1 remains comparatively less accessible. This suggests that the current cut-based strategy is sufficiently effective for BP2 while BP1 would likely benefit from further optimization (or multivariate analysis techniques).

%Figure \ref{fig:SS_vs_L} shows how the significance changes as the integrated luminosity increases. As generally expected, collecting more data increases the number of signal events, and this leads to a higher significance. The upward behavior of the curves shows that the chosen cuts can give more meaningful results at higher luminosities. In the figure, the BP2 curve stays above the BP1 curve. The main reason for this is that BP2 has a larger cross section and leaves a higher effective number of signal events after the final selections. In other words, for the same luminosity, more selected signal events are obtained for BP2. This causes the significance curve to rise faster. One of the most important results that can be seen from the figure is that the BP2 scenario becomes much stronger in terms of observability at high luminosities of the order of the HL-LHC. In contrast, the BP1 curve increases more slowly. This suggests that not only more data, but also more advanced selection strategies or multivariate analysis methods, may be needed, especially for BP1.

\begin{table}[h!]
  \centering
  \renewcommand{\arraystretch}{1.15}

  \begin{minipage}{\textwidth}
    \centering
    \setlength{\tabcolsep}{4pt}
    \begin{tabular}{lcccccc}
      \toprule
      Cuts &
      \makecell{Signal\\(BP1)} &
      \makecell{Signal\\(BP2)} &
      $t\bar{t}$ &
      $t\bar tV$ &
      $tW$ &
      $VV$ 
      %& $Z\gamma$ 
      \\
      \midrule
      Before Cuts &
      65.46 & 144.18 &
      $1.53 \times 10^{8}$ & $5.12 \times 10^{4}$ & $1.37 \times 10^{7}$ &
      $9.13 \times 10^{6}$ 
      %& 4.125 
      \\
      Preselection &
      29.880 & 78.686 &
      $1.00 \times 10^{6}$ & 295.741 & $4.39 \times 10^{5}$ &
      $1.89 \times 10^{6}$ 
      %& 2.029 
      \\
      $M(l_1,l_2) > 1250~\mathrm{GeV}$ &
      29.736 & 78.271 &
      0 & 0.180 & 0 & 243.560 
      %& 0.0112 
      \\
      \bottomrule
    \end{tabular}
  \end{minipage}

  \vspace{0.6cm}

  \begin{minipage}{\textwidth}
    \centering
    \setlength{\tabcolsep}{6pt}
    \begin{tabular}{lccccc}
      \toprule
      Cuts &
      \makecell{Total\\background} &
      \makecell{$\epsilon_S$\\(BP1)} &
      \makecell{$\epsilon_S$\\(BP2)} &
      \makecell{$\mathcal{Z}$\\(BP1)} &
      \makecell{$\mathcal{Z}$\\(BP2)} \\
      \midrule
      Before Cuts &
      $1.76 \times 10^{8}$ &
      1.000 & 1.000 &
      {0.00493} & {0.01087} \\

      Preselection &
      $3.33 \times 10^{6}$ &
      0.456 & 0.546 &
      0.01637 & 0.04310 \\

      $M(l_1,l_2) > 1250~\mathrm{GeV}$ &
      243.751 &
      0.454 & 0.543 &
      {1.799} & {4.362} \\
      \bottomrule
    \end{tabular}
  \end{minipage}

  \caption{As in Table~\ref{tab:cutflow} with $\mathcal{L}=3~\mathrm{ab}^{-1}$.}
  \label{tab:cutflow2}
\end{table}

%\begin{table}[h!]
%  \centering
%  \resizebox{\textwidth}{!}{%
%  \begin{tabular}{lcccccccccccc}
%    \toprule
%    Cuts &
%    Signal (BP2) & Signal (BP1) &
%    $t\bar{t}$ & $t\bar tV$ & $tW$ & $VV$ & $Z\gamma$ &
%    Total background &
%    $\epsilon_S^{\mathrm{BP2}}$ & $\epsilon_S^{\mathrm{BP1}}$ &
%    $\mathcal{Z}^{\mathrm{BP2}}$ & $\mathcal{Z}^{\mathrm{BP1}}$ \\
%    \midrule
%    Before Cuts &
%    144.18 & 65.46 &
%    $1.53 \times 10^{8}$ & $5.12 \times 10^{4}$ & $1.37 \times 10^{7}$ & $9.13 \times 10^{6}$ & 4.1246 &
%    $1.76 \times 10^{8}$ &
%    1.000000000 & 1.000000 &
%    {0.010866975} & {0.004933781} \\
%    Preselection &
%    78.6864753 & 29.8795 &
%    $1.00 \times 10^{6}$ & 295.741 & $4.39 \times 10^{5}$ & $1.89 \times 10^{6}$ & 2.0288 &
%    $3.33 \times 10^{6}$ &
%    0.545751667 & 0.456455 &
%    0.043102785 & 0.016367431 \\
%    $M(l_1,l_2) > 1250~\mathrm{GeV}$ &
%    78.2707563 & 29.7356414 &
%    0 & 0.179836266 & 0 & 243.55968 & 0.011205145 &
%    243.7507214 &
%    0.542868333 & 0.454256667 &
%    {4.361713254} & {1.798673796} \\
%    \bottomrule
%  \end{tabular}%
%  }
%  \caption{Cutflow at $\sqrt{s}=14$~TeV with $\mathcal{L}=3 \;\mathrm{ab}^{-1}$. }
%  \label{tab:cutflow2}
%\end{table}

Table \ref{tab:cutflow2} shows the results of the same analysis chain at the HL-LHC, i.e., for $\mathcal{L}=3~\mathrm{ab}^{-1}$. Here, the number of remaining signal events is about $78.27$ for BP2 and about $29.74$ for BP1. The total background is about $243.75$ events. The dominant part of this background still comes from the $VV$ process. Under these conditions, the significance reaches $\mathcal{Z}^{\mathrm{BP2}} \simeq 4.36$ for BP2. This result shows that BP2 comes quite close to the discovery threshold at the HL-LHC. In contrast, for BP1, the result is $\mathcal{Z}^{\mathrm{BP1}} \simeq 1.80$. This value shows that BP1 does not yet reach a strong level of observability using the same analysis at the HL-LHC. %Therefore, Table \ref{tab:cutflow2} shows that even with a simple and physically well-motivated cut-based analysis, the BP2 scenario looks quite promising at high luminosity. However, reaching the same level of sensitivity is more difficult for BP1. For this reason, more advanced analysis techniques may be needed for the BP1 scenario, such as additional kinematic variables, better optimized mass windows, or multivariate classification methods.

\section{Conclusions}

This study highlighted the potential of mono-$Z^{\prime}$ signatures at the (HL-)LHC, in leptonic final states involving electrons and muons, to explore the extended particle content of the BLSSM-IS. In fact, in such a SUSY scenario, the $Z'$ acts as efficient  portal to an extended Higgs sector which, in turn, enables the (resonant) pair production of two kinds of LSPs, a neutralino (a spin-1/2 object) or a sneutrino (a spin-0 object), via a $B-L$ Higgs boson, $h'$. Despite the different structure of the $h' {\tilde\chi}_1^0 {\tilde\chi}_1^0$   and $h' {\tilde\nu}_R {\tilde\nu}_R$  interactions, the fact that these are produced through a spin-0 object (i.e., the $h'$)  implies that the whole process is insensitive to the DM nature, hence, it can be exploited for the extraction of signals of DM whichever the candidate for the latter is. In fact, by selecting two BPs over the BLSSM-IS parameter space, wherein the neutralino and sneutrino masses are very similar (and so are the $Z'$ mass values), we have been able to show  that a variety of observables can reveal the DM presence, following a realistic MC simulation down to the detector level. We have been able to do so for the case of the HL-LHC through a dedicated selection which is insensitive to the masses of the $Z'$ and $h'$ bosons, in turn implying that the latter need not be measured to entertain the analysis that we have advocated here. However, we also recognize that sensitivity is better for the neutralino case than for the sneutrino one, so that some adjustments to the latter are needed to acquire a comparable sensitivity to both DM candidates.

\section*{Acknowledgements}

The work of SK is partially supported by the Science, Technology \& Innovation Funding Authority
(STDF) under grant number 48173. 
SM is supported in part through the NExT Institute and the Science and Technology Facilities Council (STFC) Consolidated Grant ST/X000583/1.

\providecommand{\href}[2]{#2}\begingroup\raggedright\endgroup

%\clearpage
%\newpage
%\bibliographystyle{JHEP}
%\bibliography{ISS.bib}

\end{document}